\documentclass[sigconf]{acmart}

\usepackage{xcolor}
\usepackage{multirow}
\usepackage{balance}
\usepackage{subcaption}
\usepackage{xcolor}

\AtBeginDocument{%
  \providecommand\BibTeX{{%
    \normalfont B\kern-0.5em{\scshape i\kern-0.25em b}\kern-0.8em\TeX}}}

\setcopyright{none}
\renewcommand\footnotetextcopyrightpermission[1]{}
\begin{document}

\title{No computation without representation: \\Avoiding data and
algorithm biases through diversity}

\author{C. Kuhlman}
\affiliation{%
  \institution{Worcester Polytechnic Institute}}

\author{L. Jackson}
\affiliation{%
  \institution{Howard University}
}

\author{R. Chunara}
\affiliation{%
  \institution{New York University}
}


\begin{abstract}
The emergence and growth of research on issues of ethics in Artificial Intelligence, and in particular algorithmic fairness, has roots in an essential observation that structural inequalities in our society are reflected in the data used to train predictive models and in the design of objective functions. While  
research aiming to mitigate these issues is inherently interdisciplinary, the design of unbiased algorithms and fair socio-technical systems are key desired outcomes which depend on practitioners from the fields of data science and computing. However, these computing fields broadly also suffer from the same under-representation issues that are found in the datasets we analyze. This disconnect affects the design of both the desired outcomes and metrics by which we measure 
success. If the ethical AI research community accepts this, we tacitly endorse the status quo and contradict the goals of non-discrimination and equity which work on algorithmic fairness, accountability, and transparency  seeks to address.
Therefore, we advocate in this work for diversifying computing 
as a core priority of the field and our efforts to achieve ethical AI practices. We draw connections between the lack of diversity within academic and professional computing fields and the type and breadth of the biases encountered in datasets, machine learning models, problem formulations, and the interpretation of results. Examining the current fairness/ethics in AI literature, we highlight cases where this lack of diverse perspectives has been foundational to the inequity in the treatments of underrepresented and protected group data. We also look to other professional communities, such as in the law and health domains, where disparities have been reduced both in the educational diversity of trainees and among their professional practices. We use these lessons to develop a set of 
recommendations that provide concrete steps for the computing community to increased diversity.

\end{abstract}



\maketitle

\section{Introduction}

The pervasive use of automated technologies in our society has prompted concerns regarding the fair and ethical use of large scale demographic data sets to make decisions that impact people's lives, particularly in legally regulated domains such as criminal justice, education, housing, and healthcare \cite{angwin2016machine,barocas2016big}. Along with this new paradigm come opportunities to use data analysis that is accurate, reproducible, and transparent 
to address societal issues. Many sources of data that reflect 
disparities in social outcomes come from populations that are identified as either vulnerable or underrepresented. 
Here, vulnerable populations are defined as those lacking the social capital to represent themselves including children, incarcerated persons, students, and the economically alienated/poor \cite{ShivayogiVpop}. Underrepresented groups, by contrast, are defined as individuals derived from ethnic minority populations or gender groups that have undergone historical discrimination 
and, also, as we will highlight further in this paper, are underrepresented with respect to their participation in the technology workforce \cite{googalldiv}. Data from these vulnerable and underrepresented groups show that they continue to be subject to systemic structural biases, often manifesting at data collection, that can skew the outcome of automated decision making processes. 

Indeed, the interdisciplinary community that has recently arisen to address these biases in algorithmic design and deployment has made great strides in identifying unfairness and working to address it from a computational perspective. A current limitation of the community's work is that it has not yet succeeded in fully capturing the diverse perspectives of those populations most affected by potentially biased algorithmic systems. We argue this challenges the exact problems that much of the community targets in its research output. To address these gaps, some conversation in technology communities has centered the idea of educational pipeline development as the area in which the greatest strides in ameliorating the diversity deficit can be realized \cite{SchultzPipeline}. While we appreciate the role that recruitment of underrepresented groups plays in broadening the field, we think that this approach critically under-utilizes potential diversity resources.

Thus in this work we advocate for diversifying computing and the AI research community itself as a core priority of the field and our research efforts. We limit our discussion mainly to applications of algorithmic fairness research within the social construct of North America. We acknowledge that the these challenges extend beyond the borders of the U.S., but given that us authors are based in the U.S., and the U.S. represents a relevant test case for diversity, this is most appropriate, and the main themes from this discussion are applicable and can be extended to other places.
To encourage and facilitate discussion and innovation around this goal we make the following contributions:

\begin{enumerate}
    \item We make clear the connection between the lack of diversity of communities represented in datasets and the type and breadth of the biases encountered in our data analysis, and the interpretation thereof.
   \item We highlight recent research from the ethics in AI community which illustrates cases where a lack of diversity of perspectives may have been a critical factor in the design of models and methods which suffer from unfair bias against protected groups.
    \item We identify positive efforts of the ethics in AI community with respect to the diversity deficit in computer science, while making recommendations that are informed by the best practices of fields external to computing.
\end{enumerate}
 

\section{Why Improving Diversity is Essential to the Ethics in AI Community}
\label{sec.ineq}

Bias in data and algorithms are critical issues, and efforts to address these are essential as computing researchers and practitioners design models and algorithms that are being deployed in ever more real-world scenarios.
Much scholarship within the ethics in AI community addresses unfair practices against members of vulnerable or underrepresented groups, including the explicit use of protected data attributes such as age, gender, or race or ethnicity, as well as indirect discrimination that occurs when group status is exploited inadvertently \cite{feldman2015certifying}. 
%

Bias in data may occur when there is unequal representation of protected groups.
Algorithms then trained on datasets encoding such biases can result in biased performance across groups \cite{caliskan2017semantics,chen2018my}. Additionally, even when datasets are equally representative of groups, biases in objective functions, for example optimizing for an outcome that can be driven by features of protected classes, can also result in unfair outcomes \cite{obermeyer2019dissecting}. However, even if these two issues are addressed, there certainly are other systematic issues that can pervade. Here we formally articulate the reason that systematic issues 
critically impact the work of detecting and mitigating unfair bias in algorithmic systems. 

%
Existing research has proposed many statistical ``fairness'' criteria. To a first approximation, most of these criteria fall into three different categories defined along the lines of different (conditional) independence between the random variables of the sensitive attribute $A$, the target variable $Y$, and the classifier or score $R$; independence, separation and sufficiency \cite{barocas-hardt-narayanan}. Accordingly, being based on $A$, $Y$ and $R$, these notions do not incorporate any context that may result in or perpetuate such inequalities. To further illustrate this, we examine a popular notion of discrimination defined as \emph{statistical parity} \cite{dwork2012fairness}, also referred to as disparate impact \cite{feldman2015certifying}. This notion requires that a certain group-conditional beneficial outcome rate should be the same for groups of interest. Formally, bias given as the following, should be minimized:
\begin{equation}
  |~ P(\hat{Y} = 1 | A = 1) - P(\hat{Y} = 1 | A = 0) ~|
    \label{eq:bias}
\end{equation} 
where $\hat{Y}$ is the predictor, representing $\hat{Y} : X \rightarrow Y$ a random variable that depends on $A$, $X$ and $U$. Here $A$ represents the \emph{group status} associated with an individual, defined by some \emph{protected attributes} which must not be discriminated against. $X$ represents other observable attributes of any particular individual, $U$ the set of attributes which are not observed, and $Y$ as above, the outcome to be predicted, e.g. by a machine learning algorithm. While the omission of context can be considered a limitation of the above notions, there is a possibility that by doing so, this may remove attention from, or camouflage the (broader/multilevel/structural) \emph{causes} of such inequities which can hinder their pursuit and limit sustainable equity. Following this, we formally identify the challenge of \emph{structural inequality} \cite{stolte1977structural}, and define it in line with this notation.




\vspace{2mm}

\noindent \textbf{Definition.} \textit{Structural inequality is a condition where one category of people are attributed an unequal status in relation to other categories of people, and this relationship is perpetuated and reinforced by a confluence of unequal relations in roles, functions, decisions, rights, and opportunities. Therefore, if a class $A$ = 1 is subject to a structural inequality, that would mean that $P(Y)$ is confounded, and even if $A$ is represented and
statistical parity holds (equation \ref{eq:bias} is equal to zero), the measure of bias represented by this 
formulation may not be meaningful to the full extent.}

\subsection{Impact of Structural Inequality on Algorithmic Fairness Analysis}
\label{sec.methods}

\vspace{2mm}
A structural challenge could occur in many real-world situations. Consider an example from the healthcare domain. Say getting effective treatment for a particular condition is the positive outcome $\hat{Y}=1$. Even if the probability that a patient gets treated for a particular condition is equal across all groups with attribute $A$, and there is data for all groups $A$ in the considered dataset, there could still exist an unobserved \textit{confounder}, $U$, that impacts the outcomes for groups of patients. For instance, in this case the confounder could be access to treatment due to lower levels of healthcare provider trust in particular groups of patients' use of pain medication \cite{burgess2011addressing}. 


\begin{figure}[t]
    \centering
    \includegraphics[scale = 1]{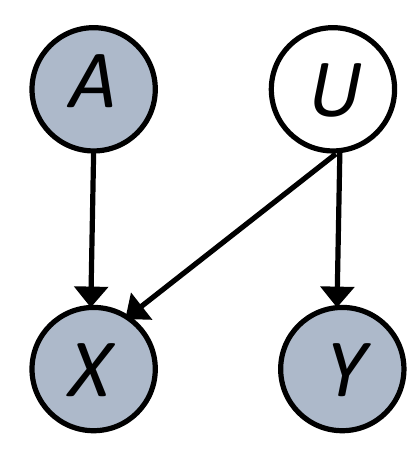}
    \caption{In a simple case, a ``confounding'' effect can be represented by $U$, which affects both $X$ and $Y$, e.g. access to healthcare which may affect groups (independent of attributes $A$ and also the outcome label for treatment. Illustrating structural inequities, including their variables and pathways need to be further identified to be accounted for in such models.}
    \label{fig:causal_diagram}
\end{figure}

\begin{table*}[t]
\centering
\begin{tabular}{l|p{2.75cm}|p{3.5cm}|p{1cm}|p{1cm}} 
\hline
\multirow{2}{*}{\textbf{Dataset Description}}&\multicolumn{4}{c}{\textbf{Sensitive Attribute}}\\
\cline{2-5}
&\textbf{race}&\textbf{gender}&\textbf{age}&\textbf{other}\\
\hline
Adult: U.S. Census income data \cite{uci}.&\cite{kearns2019empirical,friedler2019comparative,noriega2018active,oneto2019taking}&\cite{kearns2019empirical,celis2019classification,albarghouthi2019fairness,friedler2019comparative,oneto2019taking,ali2019loss,mcnamara2019costs,cardoso2019framework}&\cite{kearns2019empirical}&\cite{kearns2019empirical,albarghouthi2019fairness}\\ 
Common Crawl: Occupation biographies \cite{crawlcommon}.&&\cite{de2019bias} &&\\
Comm. \& Crime: U.S. Census and crime data \cite{uci}. &\cite{kearns2019empirical,heidari2018moral}&&&\\
COMPAS: recidivism risk assessment data \cite{angwin2016machine}. &\cite{canetti2019soft,celis2019classification,friedler2019comparative,cardoso2019framework,goel2019crowdsourcing,mcnamara2019costs}&\cite{friedler2019comparative}&&\\
Dutch census income data \cite{calders2010three}. &&\cite{cardoso2019framework}&&\\
FICO: redit scores from TransUnion \cite{hardt2016equality}.&\cite{milli2019social}&&&\\
German Credit: creditworthiness dataset \cite{uci}.&&\cite{celis2019classification}&\cite{friedler2019comparative,noriega2018active}&\\
Health Risk: proprietary scores organization\cite{obermeyer2019dissecting}.&\cite{obermeyer2019dissecting}&&&\\
Heart Health Prediction \cite{uci}. &&&\cite{noriega2018active}&\\
HMDA: Home Mortgage Disclosure Act data \cite{bureau2014using}. &\cite{chen2019fairness}&&&\\
IHDP: Infant Health and Development Program \cite{brooks1992effects}.&\cite{madras2019fairness}&&&\\
Justice: court processing data for felony defendants \cite{united1998state}.&\cite{green2019disparate}&&&\\
LSAC: National Longitudinal Bar Passage Study \cite{wightman1998lsac}.&\cite{kearns2019empirical}&\cite{kearns2019empirical}&\cite{kearns2019empirical}&\\
LSAT: Law school admission test scores and grades \cite{wachter2017counterfactual}.&\cite{russell2019efficient}&&&\\
MEPS: Medical Expenditure Panel Survey \cite{meps}&\cite{coston2019fair}&&\cite{coston2019fair}&\\
Mexican household survey \cite{ibarraran2017conditional}.&&&\cite{noriega2018active}&\\
Mobile Money Loan Approval in East Africa \cite{speakman2018three}&&&\cite{coston2019fair}&\\
PPB: Pilot Parliaments Benchmark \cite{buolamwini2018gender}.&\cite{amini2019uncovering,kim2019multiaccuracy,raji2019actionable}&\cite{amini2019uncovering,kim2019multiaccuracy,raji2019actionable}&&\\
Ricci v Stefano U.S. Supreme Court case \cite{ricci}.&\cite{friedler2019comparative}&&&\\
Stanford Medicine Research Data Repository \cite{lowe2009stride}. &\cite{pfohl2019creating} &\cite{pfohl2019creating}&\cite{pfohl2019creating}&\\
Student achievement in secondary education \cite{uci}.&&\cite{kearns2019empirical}&\cite{kearns2019empirical}&\cite{kearns2019empirical}\\
THEOP: Texas Higher Education \cite{tienda2011texas}.&\cite{borgs2019algorithmic}&&&\\
\hline
\end{tabular}
\caption{Datasets and sensitive data attributes targeted for evaluating fairness mitigation models in FAT and AIES 2019 papers.}
\label{tab.datasets}
\end{table*}

Figure \ref{fig:causal_diagram} uses a causal graph to illustrate such a scenario. The use of causal frameworks to understand the unfair impact of such confounders on potentially biased prediction has been proposed \cite{kusner2017counterfactual}, along with methods to uncover the interactions between unobserved variables and outcomes. For example, Kusner et al. \cite{kusner2017counterfactual} use a single confounder in an accident prediction model, and Kannan et al. \cite{kannan2019downstream} evaluate a hiring problem. However we know that for such complex real world interactions, there are many confounders likely to have an impact on outcomes.

What makes structural inequities more challenging than a typical confounder is found in the definition above, wherein a structural inequality is ``perpetuated and reinforced by a confluence of unequal relations in roles, functions, decisions, rights, and opportunities'' indicating that the confounder could still affect $P(Y|X)$ in an unknown/addressable way at any given point in time. In other words, by nature, the structural inequality is of an encompassing magnitude and difficult to quantify. Identifying a single variable $U$ based on this situation is not straightforward and might not fit into the simple (yet robust) paradigms often considered. 

Structural inequality may influence interactions throughout the causal graph. For instance in the hiring problem considered in \cite{kannan2019downstream} it is assumed that an employer at the end of a hiring pipeline is rational -- that it computes and makes a decision based on a posterior distribution and all necessary data is available for this. However we know that such decisions come down to the judgments of human analysts, whose  decision making is impacted by structural inequalities through their own implicit bias. Several studies have shown biases in hiring practices continue over time and have not shown any sign of decrease, despite the availability of information and policies which promote equal opportunity \cite{quillian2017meta}.

To continue our healthcare example, structural inequities may manifest through many different mechanisms. Studies have demonstrated the impact of social deprivation on health outcomes and have suggested multiple pathways that may contribute to adverse outcomes \cite{CattellSocCapHealth}. For example, patient-related health beliefs and behavior, as well as access to care through delayed presentation or access to medical services \cite{PrenticsDelayed}. Moreover, the decision to order tests can be affected by human judgments (if doctors are biased against $A$, they may be less likely to be treated (which can be referred to as the \textit{selective labels problem} \cite{lakkaraju2017selective}). However the decision to treat conditional on test results has not been shown or suggested \textit{after} testing \cite{mullainathan2019test}. Each of these occurrences can generate adverse outcomes at the individual level and also lead to structural inequities over time if there is sufficient penetrance of the described behaviors.

In sum, these examples serve to highlight how the specific issues at hand, in our example here healthcare outcome disparities, are complex and may require further domain insight or awareness in order to fully develop a given problem statement and solution formulation. Therefore in addition to the valuable approaches to fairness mitigation proposed in the recent literature, we feel it is important to consider such problems in the context of the conditions created by structural inequality.




\subsection{Impact of Structural Inequality on the Computing Community}
\label{sec.demographics}

To gain insight into the current paradigm of research into fairness and bias mitigation strategies, we consider the recent literature which focuses on the treatment of historically disadvantaged groups. Such study typically defines groups by sensitive data attributes protected by U.S. law in high impact domains \cite{barocas2016big}, including the Fair Housing Act (FHA) \cite{FHAdata} and Equal Credit Opportunity Act (ECOA) \cite{CPFBCreditdata}. The data attributes protected under these laws include age, disability, gender identity, marital status, national origin, race, recipient of public assistance, religion, and sex. 
To demonstrate the problem settings covered, we present an (incomplete) survey of recent papers from exemplary leading conferences on fairness and AI Ethics for anecdotal consideration.
Table \ref{tab.datasets} summarizes papers from the 2019 ACM Fairness, Accountability, and Transparency (FAT) Conference and 2019 AAAI/ACM conference on Artificial Intelligence, Ethics and Society (AIES). \footnote{The reader is referred to \url{https://fatconference.org/network/} for a comprehensive listing of similar venues.} Papers included are those which experimentally evaluate bias-mitigation algorithms or fairness metrics. Examining the datasets used and the protected groups targeted, we can see that the majority of analysis focuses on a narrow set of attributes, with race and gender the most prevalent sensitive attributes targeted (in 44 out of 57 experiments).

\begin{figure}[t]
\begin{subfigure}{\linewidth}
  \centering
  \includegraphics[width=\linewidth]{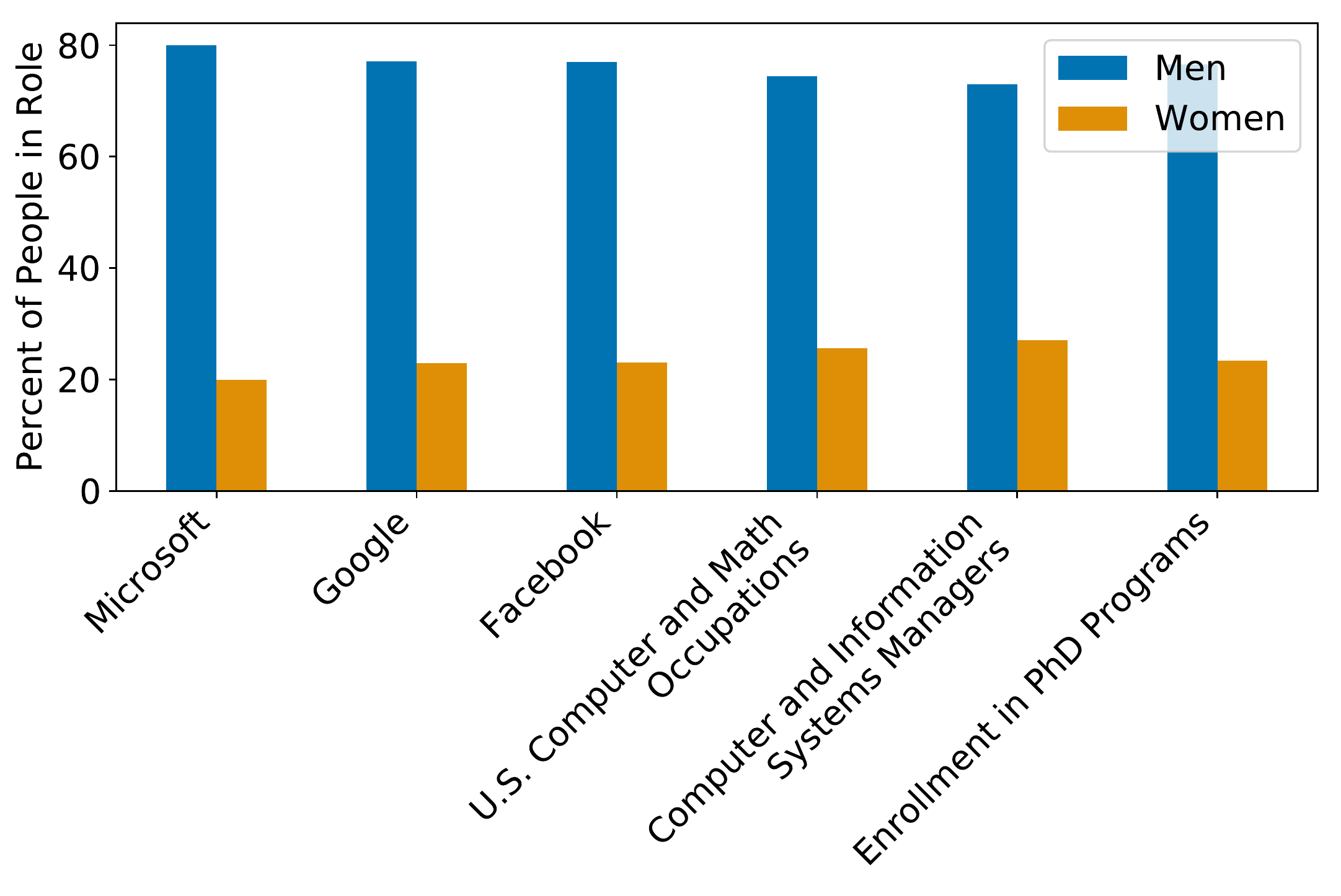}
  \caption{Gender parity values for technical employees.}
  \label{fig.sex}
\end{subfigure}
\begin{subfigure}{\linewidth}
  \centering
  \includegraphics[width=\linewidth]{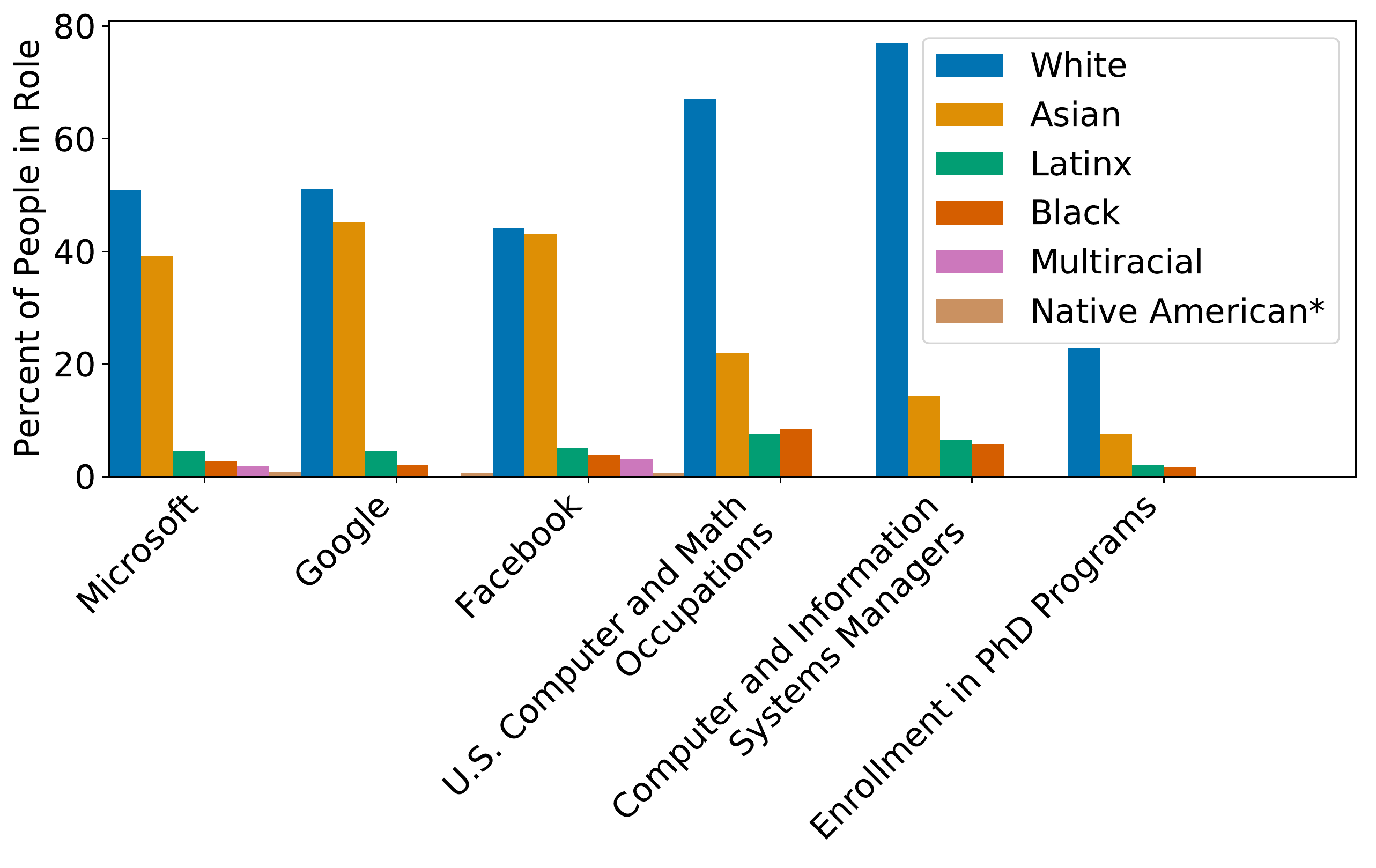}
  \caption{Breakdown of racial and ethnic groups.
*Native American includes Native Americans, Alaska Natives, Native Hawaiian and Pacific Islanders}
\label{fig.race}
\end{subfigure}
 \caption{Demographic breakdown for technical employees at top technology companies, PhD granting institutions, and computing fields in the U.S. Values for the companies are sourced from their most recent annual reports \cite{microsoft_report,google_report,facebook_report}. Educational values are from the Taulbee survey, \cite{taulbee}, and U.S. occupational data is from the Bureau of labor statistics analysis of Computer and mathematical occupations from the January 2018 Current Population Survey \cite{bls}.}
 \label{fig.dems}
\end{figure}

We present this overview of focused attention by the fairness and ethics in AI research community on the impact of structural inequality on women and racial or ethnic minorities in the United States to contrast with the inclusion of these groups in the field of computing. Unfortunately, we see stark disparity in participation of these groups in tech jobs and computing education. This disparity can be seen across computer science educational programs, research institutions, and technical jobs in industry. For instance, we refer to the Taulbee survey \cite{taulbee}, which has been conducted by the Computing Resource Association (CRA) annually since 1974. In the latest 2018 survey, across 169 PhD granting programs in the U.S. and Canada we see huge gender imbalances in computer science (77.7\% male) and computer engineering (80.7\% male) (Figure \ref{fig.race}). There is also a troubling distribution across racial or ethnic groups, with white students making up 22.9\% of enrolled students and black and latinx students making up only 2.0\% and 1.7\% respectively (Figure \ref{fig.race}). The survey also reveals the additional insight that 62.6\% of students attending these programs are from home countries different from where their institutions are located. This shows how stark the differences in engagement with computing are particularly within the U.S. population.
Similar disparity is present in industry as illustrated in Figure \ref{fig.dems}. We see the same gender imbalance exists worldwide in technical roles across top technology companies, and within the U.S. the same trend in racial disparities.  

We explicitly demonstrate this troubling disconnect between the subjects of the research in fairness and ethical AI, and the body of researchers and practitioners
here because the lack of needed domain insight and diverse perspectives has dire implications for the ability of the field to build on this crucial research and to responsibly implement the proposed methods in production systems. Failing to improve diversity in the computing field while advancing bias mitigation technologies is setting up for failure, leaving researchers and practitioners under-resourced to preempt sources of unfair bias in the technologies they design and build. 

For example, a recent study surveying machine learning practitioners was conducted to understand how tools enabled by machine learning can have a more positive impact on industry practice \cite{holstein2019improving}. Several gaps were identified from themes of this discussion.
Specifically, for the design of algorithmic systems, the crucial need to address biases in the humans embedded throughout the machine learning development pipeline was highlighted. Additionally, survey respondents noted their susceptibility to blind spots, in part due to the lack of diverse perspectives within their own teams as compared to the real-world users who interacted with their products once they were deployed. Improving the diversity of the workers and practitioners involved in this process could aid in ameliorating these issues, as it would directly provide an understanding of real world needs for appropriate product development.

It has been pointed out extensively how the lack of diversity leads to poor outcomes in many fields of endeavor. Examples range from evidence in hospitals that less diverse doctors and nurses leads to worse patient care \cite{cohen2002case,alsan2018does}, to management in firms \cite{dezso2012does}, scientific discovery \cite{nielsen2018making} and economic profit \cite{noland2016gender}. The marked gender and racial disparity we see in computing no doubt similarly impacts innovation and value in the development of new technologies.

%

\section{Fairness in the Literature and Possible Confounding}

To further elaborate on this premise that diverse representation could be a proactive approach to mitigating data and algorithm biases, we next identify specific ways in which greater diversity among the designers and creators of algorithmic systems would have been integral in avoiding the cited scenarios studied in a number of recent ethics in AI papers. 

\subsection{Biases in Data} A highly cited example of data bias is the Gender Shades study by Buolamwini and Gebru which highlighted disparate performance in commercial
facial recognition systems \cite{buolamwini2018gender}. This well-discussed scenario highlights how designers of an image processing algorithm may not think of all its implications on different populations, namely a skin color that is not their own, or representative of the majority of the people around them. Due to a lack of representation of both female faces and dark skinned faces in the training datasets used, prediction rates by these commercial systems suffered greatly for these groups.   

Another example of disparity due to training data is in natural language processing, where debiasing word embeddings has been a priority area of work \cite{bolukbasi2016man, garg2018word}. Historical stereotypes are reflected in corpora of text used to train these embedding models, which are then used widely as a pre-processing step for automated text analysis. There may be both passive and active ways of putting together image or text datasets for algorithm development, and in both these cases, a proactive approach to sourcing such datasets
could avoid wasting time and resources as well as potentially inflicting unfair or harmful outcomes on underrepresented groups.
Realizing that all sets of texts or images may not be free of bias and being in an anticipatory mode could help to address and resolve such issues.

\subsection{Algorithmic Bias} 
Another recent paper by Obermeyer et al. identifies a `problem formulation error', or in other words a mis-specified objective function as a source of unfair bias in an automated system. In this study, they examine a commercial algorithm that is deployed nationwide today and affecting millions of people \cite{obermeyer2019dissecting}.
They show that at the same health risk score, black patients are considerably sicker than whites due to the way the risk score is attributed to different illnesses that occur disparately. Instead of optimizing over health-related variables, a proxy label (in this case, cost) was used. Though not discussed by the authors, a diverse team may have identified this issue during the design of the risk score, prior to it being deployed and affecting the lives of millions of people.  

In some ways this work connects to the broad idea of problem formulation, which has been discussed \cite{passi2019problem}. This study, combining ethnographic fieldwork and ideas from sociology and history of science, as well as critical data studies, sought to describe the complex set of actors and activities involved in problem formulation. Broad conclusions demonstrated that problem specification and operationalization are always dynamic processes and normative considerations are rarely included. Their work thus also highlights the need for a broad range of perspectives and considerations at problem and algorithm formation time.

\subsection{Missing Labels} 
Another major issue in the AI ethics literature that is being addressed via algorithmic solutions is that of \emph{missing} data; specifically when membership labels for a protected class are unavailable \cite{chen2019fairness}. Indeed, much work on algorithmic fairness must assume that protected attributes are known \cite{kusner2017counterfactual, lakkaraju2017selective}. This is a reasonable assumption for work that builds on existing data, however the challenge of when a label needed to identify and ensure a class is represented/accounted for, reinforces the need for proactive recording of labels, which often are missing. This challenge may also directly benefit from diverse groups of people involved in dataset creation and analysis, who may be able to identify such attributes or recognize when they are not represented.

\section{Recommendations for Increasing Diversity within Computing and the Ethics in AI Community}

The recent literature has emphasized that increasing diversity is not simply a ``pipeline'' problem \cite{vdHurkLeaky}. As such, we next discuss three areas in which we see potential to enhance diversity and inclusion in computing research and education by engaging the ethics in AI community: (1) connecting to a broader network of higher education institutions, (2) including stakeholders from diverse communities in the research process, and (3) creating opportunities within our own activities to support a diverse group of future leaders. We look to examples of successes from other disciplines where structural inequality has impacted the diversity of practitioners and therefore outcomes in those the fields. As well, we believe that these recommendations will also address challenges in creating sustainable diversity in computing and beyond, through impacting challenges including: societal norms \cite{uhlmann2005constructed, bowles2005depends}, limited access \cite{ibarra1995race}, heterogeneous sourcing \cite{hunt2015diversity}, tokenism \cite{torchia2011women} and unfair treatment \cite{brooks2014investors, moss2012science}, which have all been described in diversity and representation gaps. While increasing diversity at the table doesn't automatically fix equity, it is thought to improve it at the individual and community level in part via exposure to different values and backgrounds \cite{richards2007addressing}.

In recommending these tangible steps we hope to build on the excellent work of the community to date. We feel the social relevance, interdisciplinary structure, and crucial importance of the emerging field of AI Ethics make it rich with potential for broadening participation in computing by appealing to students' interests and values. Making a positive social impact has been demonstrated to motivate non-traditional students in computer science education \citep{buckley2008socially,goldweber2013framework}. 
In addition, many research venues have already been evolving with diversity and inclusion as part of their core values, and this trend is encouraging. We note the success of affinity groups Black in AI, Women in Machine Learning, LatinX in AI, and Queer in AI at the NeurIPS, a top venue for Machine Learning, and the addition of the ``Critiquing and Rethinking Accountability, Fairness and Transparency'' CRAFT call in the 2020 FAT conference as evidence of these communities' openness to innovative solutions to the lack of diversity in traditional computing conferences.
Other considerations such as keeping cost of attending low, providing scholarships for students, and making conference material available online through open access and livestreams also serves to keep communities open to a wide audience and promote engagement. We applaud these efforts.

\subsection{Building Collaborations with Minority Serving Institutions}

The United States continues to see significant barriers to the full inclusion of underrepresented groups in technology disciplines, as detailed in Section \ref{sec.demographics}.
Affirmative action and other efforts to address pipeline problems have their limitations, as evidenced by analysis coming out of the fairness community itself \cite{KannanAfirm}.
One way to bolster the number of underrepresented perspectives in computing is for FAT to partner with minority serving institutions in the study of socio-technical algorithmic systems. Minority serving institutions in the U.S. include 108 Historically Black colleges and universities (HBCUs), 274 Hispanic Serving Institutions (HSIs), 35 Tribal Colleges and Universities (TCUs) and underrepresented Asian American and Pacific Islander Serving Institutions (AAPASIs).  While HBCUs comprise only 3\%
of America's institutions of higher education, they produce 24\%
of all bachelors' degrees earned by African-Americans \cite{DOImsi}. Within STEM disciplines they are responsible for graduating 40\%
of all African American STEM graduates \cite{Owens:HBCUSTEM}. Similarly, 40\% of Hispanic-American 
students are awarded their bachelors' degrees from HSIs \cite{DOImsi}.

When we broadly examine gender parity in science, technology, engineering and mathematics, it becomes clear that not every discipline has met with the same levels of gender success. When comparing gender parity in the workforce between medical schools, law schools and various tech industry staples, law school admission is one area where women have achieved parity in educational advancement since 2015
\cite{PisarcikLawSch}.
Interestingly, it has been minority serving/Historically Black Colleges and Universities (HBCUs) and non-traditional learner institutions that have been at the vanguard of this trend, 
due to their high enrollment numbers for female students. 
For example, in 2018 North Carolina Central University had a law school enrollment that was 66.85\% female; Atlanta John Marshall Law School was 66.21 \% female; Northeastern University was 65.76\% female; and Howard University 65.70 \% female enrollment. Female enrollment numbers at these institutions can be compared to the top five U.S. News and World Reports ranked law schools in the country whose female enrollment does not exceed 49.6\% \cite{EnjurisLSFemEnroll}. 

These statistics point to a potential solution to similar issues in computing education. 
While the tech industry and research institutions often focus on recruitment from
predominantly a small number of elite institutions, the lessons garnered from law school admissions suggests that partnering with non-traditional and minority serving institutions may be the way forward in addressing the lack of diversity in educational programs both in gender diversity and ethnic diversity. These minority serving institutions (MSIs) are the location of nearly half of the underrepresented trainees in computing and represent a potentially untapped 
resource of diverse perspectives. 
Furthermore, 
of particular relevance to the study of fairness and ethics in AI is the fact that these institutions have a robust intellectual tradition of contextualizing the lives of marginalized populations. 

Research partnerships can occur between individual researchers or as the result of organization to organization collaboration through a memorandum of understanding outlining specific exchanges and projects. We believe these efforts would represent a structural increase in the participation of underrepresented groups, bring a diversity of perspectives to bear on the design of algorithmic systems and thus directly address a goal of the ethics in AI community. Developing meaningful ongoing collaborations that can contextualize the implicit and sometimes explicit biases inherent in data is essential for this task.
 


\subsection{Prioritizing Research Collaboration Between the Ethics in AI Community and Underrepresented/Interdisciplinary Groups}

In addition to partnering with Minority Serving Educational Institutions, we see emphasizing collaborations with underrepresented and vulnerable groups themselves as a critical piece to broadening the scope of knowledge that the ethics in AI community has to draw upon. Recent scholarship in design-based research considers race and power dynamics between researchers and researched communities \cite{vakil2016rethinking,vakil2019racial}, and can provide guidance on designing interventions which have meaningful impacts. For instance, it has been observed that traditional eurocentric epistemologies in research communities are often disconnected from the cultural practices and ways of knowing of underrepresented and vulnerable communities \cite{bang2010cultural}. Only working closely with these communities can we begin to incorporate this knowledge into our problem designs. Having a diversity of perspectives can support the development of culturally responsive computing technologies and educational pedagogy which take a proactive inclusive approach considering intersectionality, innovations, and technosocial activism, rather than one that requires accommodations after the fact for communities left out at the development phase \cite{burgstahler2011universal,scott2015culturally}.
Engagement can happen at all stages of the data analysis pipeline, and may be particularly important during the collection, analysis, and interpretation of data from their communities. For instance, recent work \cite{JacksonVulnerable} illustrates three robust case studies for collaborations that data scientists can have with underrepresented communities, including biomedical applications for improving the health of under served populations. These projects came out of proactive collaborations with members of these communities and suggest that underrepresented and indigenous communities are 
not only interested in being the subjects of research, or the passive recipients of derived knowledge about their own communities.

The interdisciplinary nature of the ethics in AI community inherently facilitates collaboration between experts in different fields. In particular, 
collaboration with social scientists who have been assessing mechanisms for structural inequities has much to offer the field. Though the work by social scientists may not be performed in a quantitative manner, this provides an opportunity for collaboration with quantitative communities interested in fairness. In general, as identifying and quantifying sources of \textit{Structural inequality} is a complex task, it is potentially an area of interdisciplinary synergy, between quantitative scientists and the rich literature in the social sciences that already exists. This topic has been explored from multiple angles, including from the fields of education, political science, sociology, health and urban studies.

For example, a study on criminal justice algorithms might be greatly enriched by including social scientists from the communities most adversely affected by biases in the analysis of judicial data. This recommendation also can help to complete the communion loop of data findings back to those under served communities that are all to often left disconnected from the analytic fruits of their data. Creating collaborations between researchers interested in AI ethics, underrepresented data scientists, and complimentary domain expert thought leaders in these communities can lead to more robust insights into how to prevent algorithmic inequalities. Ways of accomplishing such collaboration include the example of community advisory boards that many biomedical research institutions employ to formalize academic–community partnerships \cite{newman2011peer}, as well as research-based open houses that invite the community to learn about current research projects and provide opportunities to participate in research \cite{lachney2017computational}. 

A potential outcome of these strengthened ties of research collaboration could be to increase the number of interested trainees who enter the field. This could also serve as a research idea generator where researchers use community input to define those research areas that represent their most immediate needs. Through a process of collaboration, community activists can amplify the algorithmic messages identified through careful ethics in AI research. 




\subsection{Providing Enhanced Mentorship to Trainees at Ethics in AI Research Conferences}

A barrier to full and diverse participation in the 
computing research community can revolve around the lack of onsite mentorship from senior researchers in communal spaces. Many students from diverse communities have shallower professional networks than other students and this can inhibit their introduction and advancement in research. This has a potentially pervasive effect on recruitment and retention efforts,
in opposition to indications that black and latinx students show higher interest in learning computer science \cite{googalldiv}.  Networking not only serves as an information dissemination tool, but also as a critical skill for trainees to develop for later career advancement. This can be especially challenging for trainees from underrepresented groups during networking-intense events like conferences. One ongoing concern is that underrepresented students are often less likely to have robust senior mentoring networks and strong ties to industry or research partnerships. This means that the ability to get recommendations to move forward, the prestige of those recommendations, and the ability to ask for introductions are truncated. 

One way to increase and retain diverse participants within 
the computing research community is to use volunteer research mentors at conferences who can serve as a bridge between those underrepresented trainees and early career scientists who may not be well integrated into the community and more senior participants. Pairing these individuals with those more senior researchers can rapidly expand the networks of newcomers, thereby increasing the likelihood that they will be able to make long-term contributions to the field. 

One example of a robust mentoring network exists through the Society for Molecular Biology and Evolution (SMBE). SMBE has developed a mentoring program to pair trainees and early career scientists with established researchers. This is particularly effective because their society
has an ethos of inclusiveness and contributing to the next generation of science and technologists. While their mentorship program is not limited to individuals from underrepresented groups, they make efforts to match trainees to mentors based on their merit, area of research interest, languages spoken, and geographical locations \cite{SMBEMentoring}. Trainees communicate with their conference mentors prior to the start of the meetings, giving both trainees and mentors a chance to learn about each other's research interests and career goals. During the conference, trainees and SMBE mentors are invited to have dinner together, meet up during the breaks and check in on research talks. This interaction creates an interpreted meeting that immediately connects neophytes with institutional knowledge. This link between well-networked researchers and those in need of connections can also facilitate retention of trainees and promote enthusiasm for the discipline amongst a broaden cohort of participants.

An important feature of the SMBE mentoring program is the inclusion of travel support for trainees whose characteristics will broaden the capacity of the organization to reach diversity goals \cite{SMBEMentoring}. Travel support for vulnerable and underrepresented trainees and early career scientists is a necessary investment in diversity and inclusion for computing communities.   
While merit only based awards are an excellent way to incentivize groundbreaking research, we observe that these programs often play into the pre-existing resource imbalances that favor trainees at elite institutions where resources are less constrained than they are at most MSIs. This means that institutions that train the large portions of underrepresented graduates have the least likely path to participation in conferences and workshops, while having the greatest financial barriers to entry into these intellectually rich spaces.

In Poverty and Power, Royce asserts that social networks are particularly important for underrepresented or marginalized populations because the ties binding people together and connecting individuals to organizations can: 
\begin{quote}
    "\dots channel information, convey cultural messages, create social solidarities, forge expectations and obligations, facilitate the enforcement of social norms, engender relations of mutual trust, serve as sources of social support, and operate as conduits of power and influence. In the performance of these functions, furthermore, social networks shape the distribution of resources and opportunities, advantaging some and disadvantaging others."
    \cite{Royce:PovPower}
\end{quote}
This recitation of the comprehensive utility of social networks supports our assessment that mentoring, a modality to build robust social networks, can serve as a necessary tool to build ethics in AI community diversity amongst trainees and early career scientist from underrepresented populations. 

The use of affinity and mentoring workshop programs, such as those mentioned previously and Broadening Participation Workshops associated with conferences in a variety of computing sub-disciplines, have the secondary effect of building a new cadre of research leaders who can continue to invest their intellectual and service efforts towards the betterment of the field. Investment in trainees and early career researchers is an essential part of organizational capacity building. The ethics in AI community is a relatively new research community that could see substantial benefits from the increased inclusion of trainees and early career scientists from underrepresented disciplines. 

One such workshop program is the Broadening Participation in Data Mining (BPDM) Workshop. This workshop traditionally brings together underrepresented trainees and early career scientists to increase their exposure to academic, industry and federal careers in data science. In 2019, the 7th annual BPDM workshop brought together 55 trainees and 10 mentors to Howard University, a Historically Black College/University. In the past, BPDM was associated with national or international conferences such as ACM SIGKDD or SIAM CSE. This 3 day workshop allowed participants to network with peers and senior mentors within a community of other individuals from underrepresented communities. As part of a computing tutorial exercise during the workshop, participants spent time using the COMPAS dataset to identify key themes and possible solutions to structural inequality \cite{JacksonVulnerable}. This tutorial session was important not only because it addressed fairness and ethics principles within a computer science context, but it also served to encourage participants to use computing to address issue that were of considerable concern to the social justice needs of their communities. Finally the workshop provided opportunities to create intellectual community for those participants whose experience in undergraduate, graduate and post-doctoral education is isolating by virtue of their identity.






\section{Conclusion}
Education efforts that grow representation in meaningful ways may obviate the need for much algorithmic manipulation, and also help to maintain retention efforts because no one member of an underrepresented community has to shoulder the burden of speaking for their group. This work recognizes opportunities for the ethics in AI community to increase the breadth of perspectives in computing in order to further develop our pursuit of algorithmic fairness. We have proposed three major areas within which the community can address structural inequalities: building educational collaborations with minority serving institutions, building capacity through research collaborations with community domain experts, and using educational mentoring to develop a cadre of diverse future leaders in the computing. This can be accomplished by meeting underrepresented and vulnerable communities `where they are.' This idea 
refers to both identifying the educational institutions that produce the largest number of underrepresented trainees, and using mentoring approaches to increase opportunities for trainees to actively participate in communal spaces such as
computing 
conferences.

We support an organic bottom up approach that both assists researchers in improving the fairness of algorithmic systems while empowering under served gender and ethnic communities to realize equity. This capacity building approach within research communities can also help to reinforce more equitable resources for research endeavours for collaborating partners, and increased communal activism to ameliorate structural inequalities reflected in data that the ethics in AI community works to account for. Without attempting to enact these educational initiatives, the ethics in AI community may miss out on a unique opportunity to build diverse perspective capacity, and to sustainably improve fairness, accountability and transparency in socio-technical systems through addressing structural inequalities. 





\balance
\bibliographystyle{ACM-Reference-Format}
\bibliography{acmart}





\end{document}